\documentclass[prb,twocolumn,amssymb,showpacs]{revtex4}
\usepackage{graphicx}
\usepackage{amsmath}

\begin{document}

\title{Structure of MnO nanoparticles embedded into channel-type matrices.}
\author{I. V. Golosovsky}
\affiliation{Petersburg Nuclear Physics Institute, 188300, Gatchina, St. Petersburg, Russia.}
\author{I. Mirebeau}
\affiliation{Laboratoire L\'eon Brillouin, CE-Saclay, F-91191, Gif-sur-Yvette, France.}
\author{E. Elkaim}
\affiliation{LURE, Centre Universitaire Paris-Sud, 91898, Orsay, France.}
\author{D. A. Kurdyukov and Y. A. Kumzerov}
\affiliation{A. F. Ioffe Physico-Technical Institute, 194021, St. Petersburg, Russia.}

\begin{abstract}
X-ray diffraction experiments were performed on MnO confined in mesoporous silica SBA-15 and MCM-41 matrices
with different channel diameters. The measured patterns were analyzed by profile analysis and compared to
numerical simulations of the diffraction from confined nanoparticles. From the lineshape and the specific
shift of the diffraction reflections it was shown that the embedded objects form ribbon-like structures in the
SBA-15 matrices with channels diameters of 47-87 {\AA}, and nanowire-like structures in the MCM-41 matrices
with channels diameters of 24-35 {\AA}. In the latter case the confined nanoparticles appear to be narrower
than the channel diameters. The physical reasons for the two different shapes of the confined nanoparticles are
discussed.
\end{abstract}

 \maketitle

\section{INTRODUCTION}

The study of the physical properties of confined nanomaterials has become a very active field of research
during the last decade. The reason to investigate such materials is fundamental since the confined geometry and
the influence of the surface yields unusual properties as compared with the bulk. Confinement may also result
in new applications, for example, in the field of catalysis . Since the physical properties depend on both the
size and the shape of the nanoparticles, their structural study is a crucial starting point to any detailed
investigation.

Our systematic investigation of the classical antiferromagnet MnO confined in different matrices, started with
a study of MnO embedded in a vycor-glass matrix with a random network of pores, and revealed remarkable
differences with the bulk properties \cite{MnO-PRL}. We have now performed a structural study of MnO embedded
in channel-type MCM-41 and SBA-15 matrices, known as mesoporous molecular sieves, using synchrotron radiation
and neutron diffraction. The main purpose of this work is the determination of the real shape of the confined
objects from the analysis of the diffraction patterns and their numerical simulation.

Diffraction from nanoparticles within nanochannels has many common features with diffraction from carbon
nanotubes and many publications have been devoted to its simulation in recent years. These works can be
separated in
two groups. First, the papers focusing to the description of the electron diffraction experiments at small wave
vector transfers using direct methods, as shown for example in ref. \cite{Lucas}. The direct calculations
of theoretical x-ray diffraction patterns were performed for ultra-small zeolite crystals with a complex unit
cell, which includes a large number of atoms. In these calculations one uses the reciprocal lattice numerical
integration to convolute a structure factor with an interference function \cite{Schlenker}.

Another large group of studies is related to powder diffraction from bundles of carbon nanotubes
\cite{Reznik,Rols} and to diffraction from graphite layers or thin monolayers of different adsorbates
\cite{Stephens,Kjems}. These studies are based, in turn, on the classical theory of diffraction from an ideal
two-dimensional lattice \cite{Warren}.

In our diffraction experiments we found that the lineshape from the MnO particles embedded in nanochannels of
some samples has a specific "sawtooth" profile indicative of a two-dimensional structure. To elucidate such
data a profile analysis and numerical simulation of the diffraction profiles from different nanoparticles were
performed. The lineshape of the diffraction peak and the position of its maximum give information about the
shape and dimensions of the nanoparticles.

The knowledge of the microscopic structure of the embedded MnO particles is crucial to understand their
magnetic behavior. This behavior, deduced from neutron diffraction, ESR and magnetization experiments, will be
reported in future papers \cite{neutrons,ESR}.

The paper is organized as follows. First, we show that the profile analysis of the experimental patterns can
give information about the shape of the nanoparticles embedded in channels of different diameters. Then using
numerical calculations of the diffraction patterns, evaluations of the nanoparticles dimensions are made. We
demonstrate that in the matrices with the largest channels the confined objects form thin ribbon-like
structures, whereas in the matrices with the narrowest channels MnO adopts the shape of nanowires.

\section{Experiment details.}

The experiments were performed with MnO embedded in different channel-type matrices. MCM-41 type matrices
\cite{Grun} with 24 and 35 {\AA} channel diameters (referred below as MCM) and SBA-15 type matrices \cite
{Zhao} with 47, 68 and 87 {\AA} diameters (referred below as SBA) were used. MCM-41 and SBA-15 matrices differ
by the preparation technique. Both present an amorphous silica (SiO$_{2}$) matrix with a regular hexagonal
array of parallel cylindrical nanochannels.

Their specific surface, measured by N$_2$ adsorption was between 400 and 700 m$^2$/g for SBA, and up to 960
m$^2$/g for MCM. SBA can reach higher porous diameters of the channels than MCM, but the wall roughness is also
higher. The distance between adjacent nanochannels, namely the wall thickness, is constant and of about 8-10
{\AA} in MCM \cite{Grun}, whereas it varies with the channel diameter in SBA.

The matrix powders with a grain size $\sim $ 1-2 ${\mu}m$ were prepared in the Laboratoire de Chimie Physique,
Universit\'e Paris-Sud, France \cite{Morineau-2}. All samples were filled with MnO by the "bath deposition"
method from a solution in the Ioffe Physico-Technical Institute (St Petersburg, Russia). The high specific
surface of the matrices and the good wetting of the channel walls by the liquid ensure that MnO predominantly
occupies the channel voids. This is confirmed by a consistent analysis of neutron diffraction, X-ray
diffraction, ESR and magnetization, to be published elsewhere.

Neutron diffraction experiments were carried out at the diffractometer G6-1 of the Laboratoire L\'eon Brillouin
at the Orph\'ee reactor with a neutron wavelength of 4.732 {\AA}. X-ray diffraction experiments were performed
at the beam-station WDIF 4C at LURE (Laboratoire pour l'Utilisation du Rayonnement Electromagn\'etique) with a
wavelength of 1.000 {\AA}. Measurements were carried at room temperature, keeping the Debye-Scherrer geometry.
To avoid preferred orientation effects, the powder samples sealed in thin quartz capillaries were rotated
continuously during the experiment.

\begin{figure} [t]
\includegraphics* [width=\columnwidth] {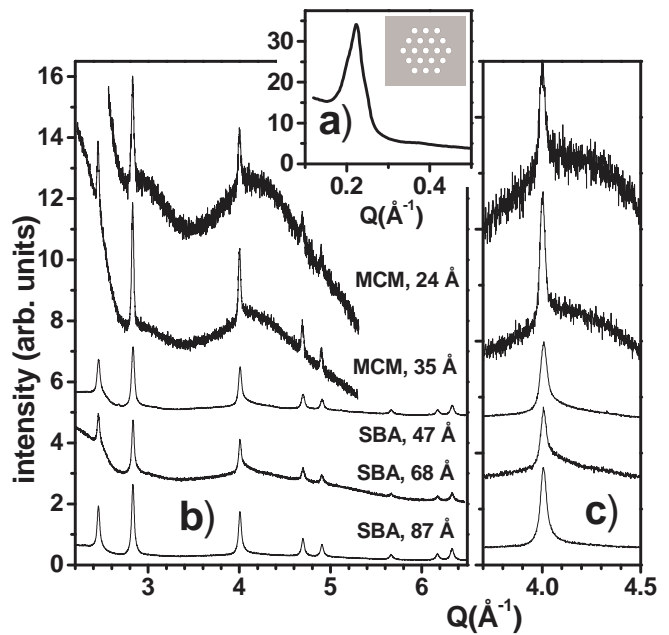}
\caption{a) Reflection (10) from the hexagonal array of nanochannels in the MCM matrix with 24 {\AA} channel
diameter. In the inset, a schematic drawing of the hexagonal superlattice. b) X-ray diffraction patterns of MnO
confined in nanochannels, normalized to the scale factors refined from the integral peak intensities. The
decrease of the signal with decreasing channel diameter is reflected in the increase of the statistical error
(noise). c) Reflection \{220\} in enlarged scale.} \label{experiment}
\end{figure}

\section{Experimental results and data treatment.}

\subsection{Diffraction.}

The honeycomb packing of the nanochannels yields an intense reflection (10) from the two-dimensional hexagonal
lattice. Due to the large periodicity the (10) reflection is observed in the low-\textit{q} region. In figure
\ref{experiment}, insert a, the (10) reflection measured by neutron diffraction from MCM with channels of 24
{\AA} diameter is shown, together with a schematic drawing of the hexagonal superlattice. The period and
coherence length calculated from the position and broadening of the (10) peak are 31.82(5) and 132(5) {\AA},
and 42.38(9) and 233(3) {\AA}, for MCM with 24 and 35 {\AA} channel diameter, respectively. This means that the
wall thickness for MCM is about 7.5 {\AA}.

The X-ray diffraction patterns measured at room temperature on the matrices with embedded MnO are shown in
figure \ref{experiment}b and in enlarged scale in figure \ref{experiment}c. In the diffraction patterns from
SBA, the so called "sawtooth" profiles are clearly seen, in contrast with patterns from MCM, where this feature
is not obvious. The asymmetrical line shape has a rapid increase on the low angle side and a long tail on the
high angle side. This profile is well known in powder diffraction patterns from carbon nanotubes, graphite
layers or different adsorbates on graphite and is associated with the two-dimensional periodicity of the layer
\cite{Kjems}.

We used approximately equal sample volumes, therefore the decrease of the signal with decreasing the channel
diameter corresponds to a decrease of the amount of the oxide confined in the channels. For example, the
intensity of the signal in the MCM with 24{\AA} channel diameter is about 25 times smaller than in the SBA with
87{\AA} diameter. One should stress that the quantity of MnO embedded in the matrix does not only depend on the
channel size, but also on some specific features of the matrix, the procedure of filling, and the wetting of
the walls by the embedded material.

It is known that the Mn oxidation process progresses on the surface \cite{Sako}. Without special precautions
MnO transforms into an amorphous state after a month. Because of the enormous surface area (corresponding to a
specific surface of 500 to 1000 m2/g), it is impossible to avoid the presence of this amorphous phase in the
present samples. We attribute the differences in the background modulations observed in MCM and SBA to
different fractions of the amorphous phase.

\subsection{Profile analysis.}

From the pioneering work of Warren \cite{Warren} it is known that diffraction from a two-dimensional lattice
has two characteristic peculiarities. Firstly, a specific "sawtooth" profile, which in the ideal case is
described by the Warren integral. Secondly, a displacement of the peak maximum from the Bragg position towards
larger diffraction angles $2\Theta$. This second effect yields an "effective" unit cell parameter which is
systematically lower than the parameter expected for a three-dimensional lattice.

\begin{figure} [t]
\includegraphics* [width=\columnwidth] {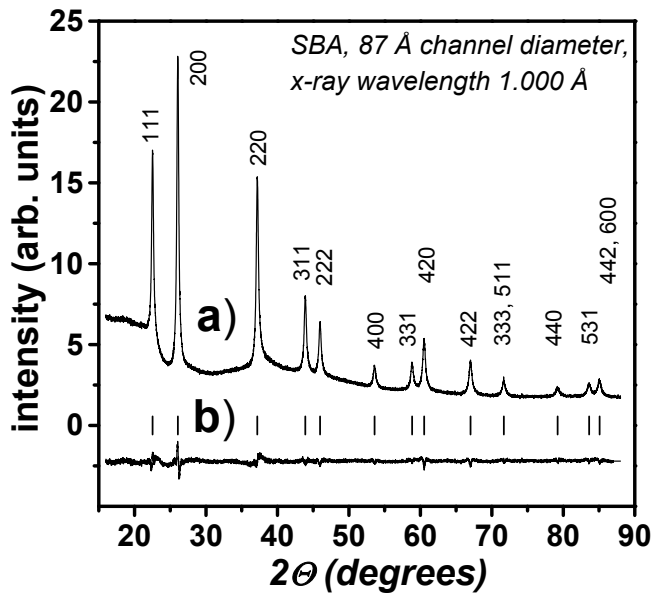}
\caption{ a) Observed X-ray diffraction pattern of MnO confined in SBA matrix with 87 {\AA} channel diameter;
b) difference (calculated - observed) pattern. Diffraction reflections are shown by vertical bars.}
\label{fullprof}
\end{figure}

\begin{figure} [t]
\includegraphics* [width=\columnwidth] {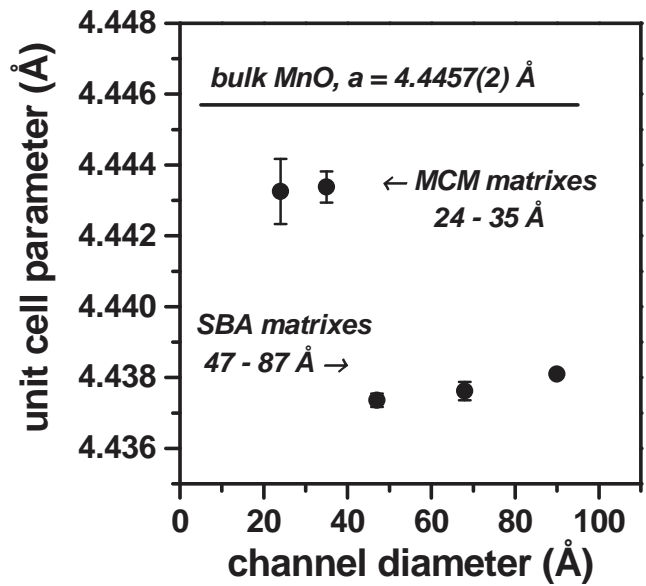}
\caption{The "effective" unit cell parameters, refined from the profile analysis.} \label{cell}
\end{figure}

The comparison of the experimental profiles with the theoretical ones, calculated by the Warren theory
developed for an ideal two-dimensional lattice, shows that the intensity of the experimental profile drops with
increasing diffraction angle much faster due to the finite thickness of the real structure. However the
position of the peak maximum does not change significantly, since it is defined by the characteristic size of
the two-dimensional lattice. Therefore, in a first step we used the "effective" lattice parameter calculated
from the positions of the maxima to evaluate the dimensions of the diffracting nanoparticles.

From the theory of diffraction from an ideal two-dimensional lattice the peak displacement is given by the
formula \cite{Warren}:

\begin{equation}
{\Delta\left({sin\left({\theta}\right)}\right) = 0.16\cdot\lambda/L} \label{eq1}
\end{equation}

\noindent where \textit{L} is the characteristic size of the two-dimensional lattice and $\lambda$ is the
incident wavelength. The modulus of the scattering vector being $q = 4\pi{sin{\Theta}}/{\lambda}$, the equation
(\ref{eq1}) can be rewritten as ${\Delta{q} = 2.01/L}$. The shift of the Bragg reflections in \textit{q}-space
will be referred below as the "Warren shift", which corresponds to some "effective" lattice parameter. The
latter, in turn, can be refined using the Rietveld method.

To perform the profile analysis with strongly asymmetric reflections, we used a specific description of the
line shape, namely, the "split" one, which is implemented in the FullProf program \cite{Fullprof}. In this
model each peak is separated into left and right parts with respect to the position defined by a regular
lattice with an "effective" lattice parameter. Each part of the peak is approximated by a separate pseudo-Voigt
function whose parameters are refined independently. The parameters describing the peak broadening due to
strain and size effects are the same for the two parts of the peak.

The proposed approximation describes the observed X-ray pattern satisfactorily (figure \ref{fullprof}). As
expected, the right part has a Lorentzian shape, while the Lorentzian contribution in the left part is much
smaller. Refining the common parameters shows that the observed peak broadening is due only to a size effect
and that the contribution from the inner strains is negligible. No preferred orientation of the nanoparticles
was detected. This means that there is no texture and that the samples are isotropic in the sense of powder
diffraction.

In figure \ref{cell} the refined lattice parameter is shown as a function of the channel diameter. As a
reference parameter we used the unit cell parameter of the bulk MnO \cite{Morosin}, displayed by a horizontal
line in figure \ref{cell}. For MCM with smaller channels of 24 {\AA} and 35 {\AA} the refined unit cell
parameter appears to be close to the bulk value in contrast with SBA with larger channels. The peak asymmetry
and the "effective" lattice constant suggest two different types of nanoparticles in MCM and SBA.

To better understand the real structure of the embedded objects we undertook a numerical simulation of the
diffraction patterns.

\section{Numerical simulation and dimensions of the diffracting objects.}

There are two different approaches to model the diffraction from complex objects: analytical methods and direct
numerical computation. We used the straightforward numerical computation of the scattering intensity from a
powder \textit{I(q)} based on the Debye formula \cite{Yang}:

\begin{equation}
I(q) \sim \frac{1}{N}\sum\limits_{i,j = 1}^N{A_{i,j}\frac{{sin(R_{i,j}q)}} {R_{i,j}q}} \label{eq2}
\end{equation}

\noindent where ${R_{i,j}}$ is the distance between the \textit{i-}th and \textit{j-}th atoms, $A_{i,j}$ is the
number of the specific distance ${R_{i,j}}$, \textit{N} is the total number of all possible distances in the
considered volume and \textit{q} is the wave vector transfer. Such a universal approach assumes the orientation
of the diffracting object to be completely random and is suitable for a diffracting object of any shape
\cite{footnote}.

Since we are interested only in the lineshape of the reflections and not in the intensities, the atomic
form-factors were not taken into account. To simplify calculations we considered only a monoatomic lattice.
This assumption is valid in the case of X-rays because of the large difference in the charges of Mn and O ions;
the Mn contribution to the diffraction pattern dominates.

We used the direct algorithm by computing of all possible specific distances ${R_{i,j}}$, then we used the
recursive procedure of quick sorting and a successive calculation of the numbers ${A_{i,j}}$. This is a rather
rapid method, and it is practically limited only by the available computer memory.

\subsection{Diffracting objects in SBA matrices.}

\begin{figure} [t]
\includegraphics* [width=\columnwidth] {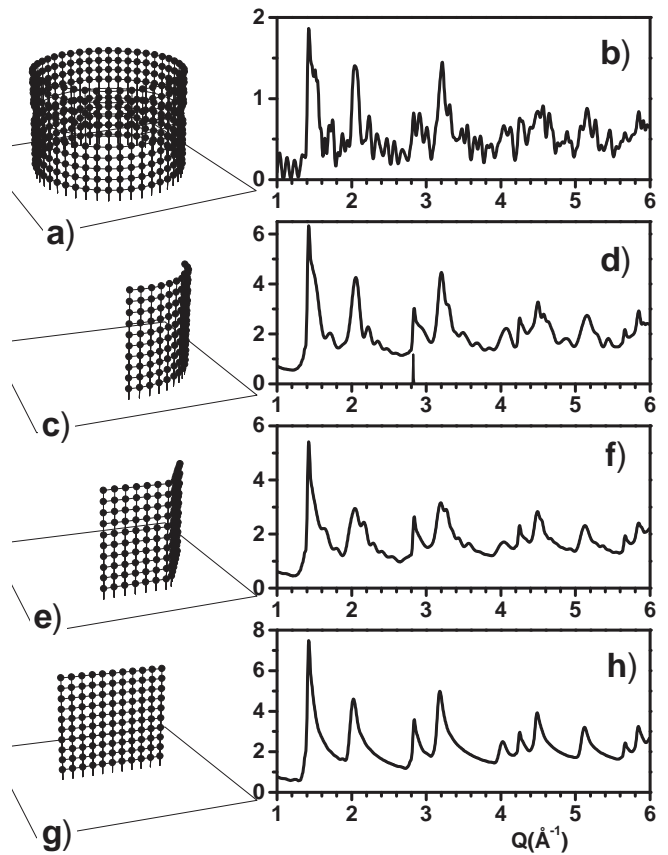}
\caption{Numerical simulation of diffraction patterns from diffracting objects of different shapes. All objects
are constructed from an ideal two-dimensional square lattice with a lattice parameter of 4.446 {\AA}. At the
left, models of the diffracting objects are shown. The instrumental resolution is shown by a vertical bar at $q
= 2.83$ {\AA}$^{-1}$ at the bottom of pattern d).} \label{simulation}
\end{figure}

\begin{figure} [t]
\includegraphics* [width=\columnwidth] {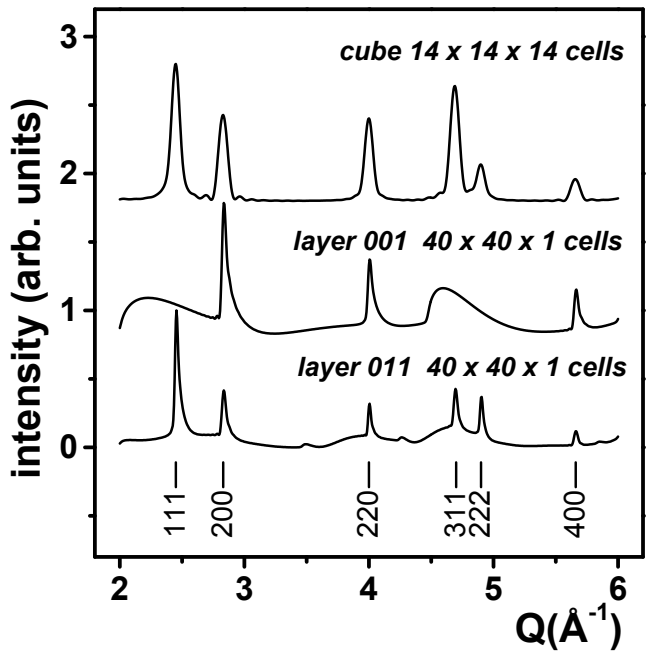}
\caption{Numerical simulation of diffraction patterns from different diffracting objects: cube and two thin
layers with different orientation with respect to the crystallographic axis. The labels "layer 001" and "layer
011" mean that the layers are perpendicular to the [001] and [011] axis, respectively. All unit cells have a
fcc lattice with a unit cell parameter of \textit{a} = 4.44 {\AA}.} \label{simulation-layers}
\end{figure}

Due to their specific "sawtooth" profile the diffracting nanoparticles in SBA are obviously thin layers. In
principle, the thin layers of MnO within the channels of SBA could crystallize in the form of cylindrical
surfaces or as flat objects. To clarify this question we performed numerical simulation of the diffraction
patterns from different fragments of an ideal two-dimensional square lattice inscribed in a cylindrical
surface, namely, a full cylinder, a limited cylindrical surface, a polygonal surface and a flat layer (figure
\ref{simulation} a, c, e and g, respectively).

To compare with the real object, we used a cylindrical surface with a diameter of 14 unit cells, which
corresponds to 62 {\AA} for the unit cell of MnO, and with a length of 40 cells (178 {\AA}) along the cylinder
axis. The primitive lattice was used instead of the fcc lattice to simplify calculations, because it does not
affect the lineshape of the reflections.

In figure \ref{simulation} the results of computer simulations are shown for a full cylinder, a limited
cylindrical surface, a polygonal surface and a flat layer. As expected the calculated diffraction profile from
the cylindrical surfaces shows significant distortions of the lineshape because of a partial lost of coherence
as compared with a flat layer. For example, the difference in the distances between adjacent atoms along and
perpendicular to the cylinder axis, which is absent in a flat lattice, leads to the peak splitting.

The high resolution of the synchrotron experiments (vertical bar at the bottom in figure \ref{simulation}d)
cannot modify the profiles. However, the profiles could be smeared by a possible distribution in the dimensions
of the diffracting objects. Our simulations show that the "sawtooth" profile transforms into a symmetric one
very rapidly with a small increase of the layer thickness. Therefore we only considered a size distribution in
the largest dimension L of the layer. Simulations using a Gaussian type distribution show that a size
distribution with $\Delta{L}/L$ up to 40 {\%} has no pronounced effect on the calculated profile. We obtain the
same result by calculating the width of the (10) reflection for different distributions of diameters in the
nanochannels, as done in ref\cite{Rols}.

By comparing the calculated profiles with the observed patterns (figure \ref{experiment}) we conclude that the
diffracting objects can be curved only slightly within the channel and that they more likely consist of
fragments of flat bands or ribbons with a width close to the channel diameter. A similar conclusion was reached
from X-ray diffraction results from carbon nanotubes which was attributed to a polygonization of the nanotubes
\cite{Reznik}. Moreover, our results are fully consistent with the X-ray structure modelling of MCM-41
matrices, where it was shown that channel shape appears to be much closer to a hexagonal prism than a cylinder
\cite{Solovyev}.

From the theory of diffraction it is well known that the peak broadening from objects of anisotropic shape
strongly depends on the angle between the scattering vector and the axes of anisotropy \cite{Langford}. The
broadening of a particular reflection is inversely proportional to the "apparent size" of the object, namely,
to its average thickness measured along the scattering vector (review \cite{Langford-review}). If an object has
a cylindrical or layered shape, its apparent size varies with the direction of the scattering vector
considered. This leads to a peak broadening, which is the same for all reflections whose scattering vectors
have the same angle with the anisotropy axis. For example, Pb embedded in a porous vycor glass crystallizes in
the form of cylinder along [111] axis, that results in the systematic narrowing of all reflections \textit{hhh}
type \cite{Pb}.

A similar effect is illustrated in figure \ref{simulation-layers} for thin layers. The profile at the top is
calculated for a cube with the dimensions of $14 \times 14 \times 14$ fcc cells as a model for an infinite
crystal. The two profiles below are calculated for two layers with the same dimensions of $40 \times 40$ cells
and a thickness of one cell, but with different orientations of the layer plane with respect to the
crystallographic axis.

The diffraction profiles from layers with different orientations significantly differ. Indeed, for the layer
with one cell thickness oriented along [100] and [010] directions, the thickness (apparent size) along all
\{111\} directions is $a\sqrt{3}$, while the apparent sizes along [100], [010] and [001] directions are of
$40a$, $40a$ and $a$, respectively. Therefore, since the peak broadening is inversely proportional to the
apparent size, the 111 reflection (sum of all overlapping reflections \{111\}) is much broader than the \{200\}
reflection (sum of two narrow 200 and 020 and one broad 002 reflection).

We do not know the real contribution to the diffraction reflections from layers with different crystallographic
orientations. Therefore, it is impossible to obtain quantitative results from the peak broadening. However the
"Warren shift" is not strongly sensitive to the peak shape; therefore, some estimations of the dimensions of
the confined objects can be made on this basis.

\subsection{Estimation of the layer dimensions in the large channels.}

Estimations of the layer dimensions can be done by comparing the relative deviation of the unit cell parameter
$\Delta{a}/a$ due to the "Warren shift" calculated for different objects based on the MnO lattice with the
experimental value.

\begin{figure} [t]
\includegraphics* [width=\columnwidth] {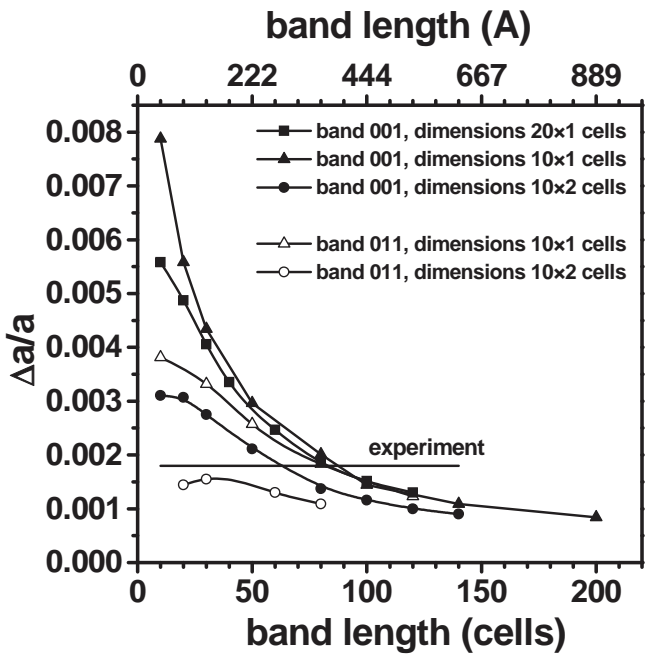}
\caption{Warren shift : deviation of the "effective" lattice parameter from the lattice parameter of an
infinite lattice, calculated for thin layers of different sizes and orientations.} \label{cell-layer}
\end{figure}

As shown in figure \ref{cell}, the effective unit cell parameters are close to each other for all objects
inside SBA. Therefore we used a mean value, shown as a horizontal line in figure \ref{cell-layer}. The
interception of this line with the calculated curves gives the estimation of the layer dimensions. The
calculations were made for layers with a fcc lattice, choosing two possible orientations of the layers with
respect to the crystallographic axis (figure \ref{cell-layer}).

Analyzing $\Delta{a}/a$ calculated for different layers (figure \ref{cell-layer}), it is shown that for layers
with a similar thickness a variation of the layer's width in the range of 20-10 cells, which corresponds to the
channel diameters of the SBA, has little effect if the layer's length is more than 80 cells. In contrast, the
layer thickness and layer orientation, which define the "apparent size", i.e. the real thickness, strongly
affect the "Warren shift". Whatever their orientation, the thickness of the layers cannot not be more than 2
cells. By comparing the calculations with the experimental value (straight line), the following rough
estimation is obtained of the layer dimensions for the nanoparticles in SBA: the layer thickness $\sim $ 1-2
cells (4-9 {\AA}), the ribbon width $\sim $ 10-20 cells ($\sim $ 44-88 {\AA}) and the ribbon length $\sim $
60-80 cells ($\sim $ 270-350 {\AA}).

In summary, MnO crystallizes inside SBA with channel diameters of 47-87 {\AA} adopting the shape of thin,
narrow and long ribbons, which are possibly slightly curved.

\begin{figure} [t]
\includegraphics* [width=\columnwidth] {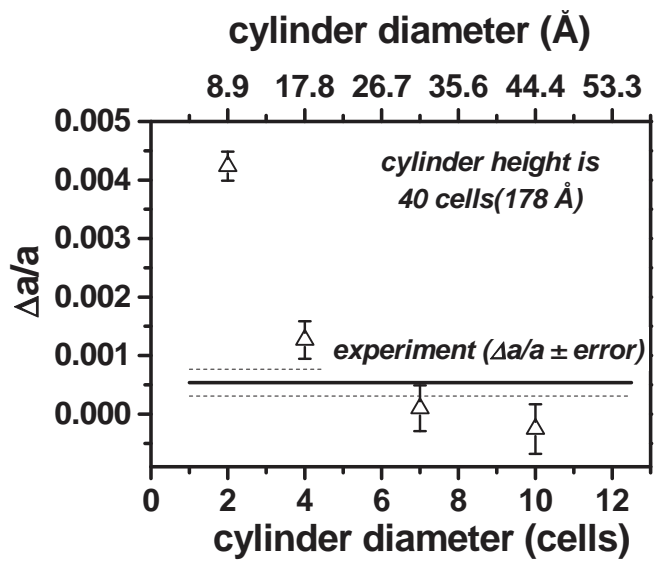}
\caption{Deviation of the "effective" lattice parameter from the parameter of an infinite lattice
$\Delta{a}/a$, calculated for thin cylinders.} \label{cell-cylinder}
\end{figure}

\subsection{Diffracting objects in MCM matrices.}

The diffraction peaks from MnO within the narrow channels of MCM do not show any obvious "sawtooth" profile
(figure \ref{experiment}). Since the channel diameters of MCM are small, 24 and 35 {\AA} (4-8 cells), it is
natural to assume that the diffracting objects have the shape of narrow cylinders. Diffraction reflections from
such objects are symmetrical, therefore their full width at half maximum (FWHM) should be a good parameter to
identify their dimensions.

Refinement using the Thompson-Cox-Hastings approximation of the lineshape with independent variation of the
Gaussian and Lorenztzian contributions \cite{Thompson} shows that the observed peak broadening is due to a
size effect only, without any contribution from inner stresses. The volume averaged sizes calculated from the
peak broadening are 168(2) {\AA} and 204(2) {\AA} for matrices with 24 {\AA} and 35 {\AA} diameters,
respectively. Because the channel diameters are much smaller, the refined values are very close to the
cylinder heights. These results are fully consistent with the results obtained by comparing the measured full
width half maximum (FWHM) and the calculated one using a Debye formula for cylinders of different dimensions.
This method is similar to the method used for nanoribbons, except that here we compare the values for FWHM's instead of
for the Warren shift.

\begin{figure} [t]
\includegraphics* [width=\columnwidth] {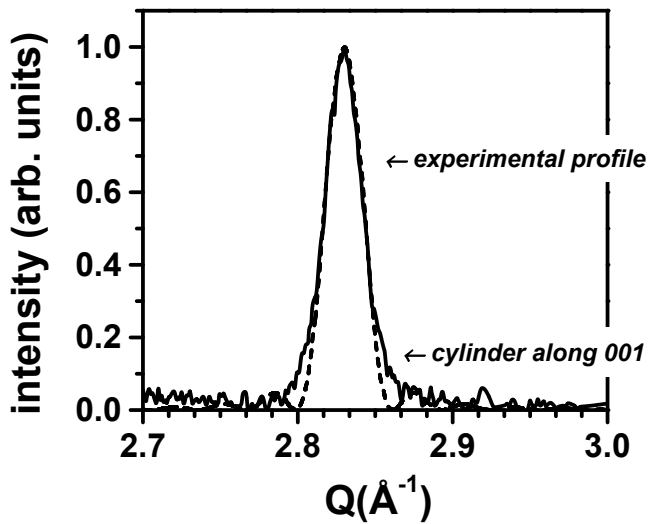}
\caption{Experimental profile of \{200\} reflection (solid line) and profile calculated for cylinder with the
dimensions of ${\oslash}{\ }5\times40$ cells aligned along [001] direction (dash line).} \label{profiles-3}
\end{figure}

Numerical simulations of the lineshape for diffracting cylinders show that the "Warren shift" also exists for
the narrow cylinders. Moreover, this shift appears to be very sensitive to the cylinder diameter and can be
used for its evaluation. The $\Delta{a}/a$ calculated for cylinders of 40 cells height with different diameters
is shown in figure \ref{cell-cylinder}. The observed shift averaged for two MCM samples is shown by a
horizontal line with a confidence range. Its intersection with the calculated curve shows that the diameter of
the cylinders are between 4-6 cells (18-27 \AA).

\begin{figure} [t]
\includegraphics* [width=\columnwidth] {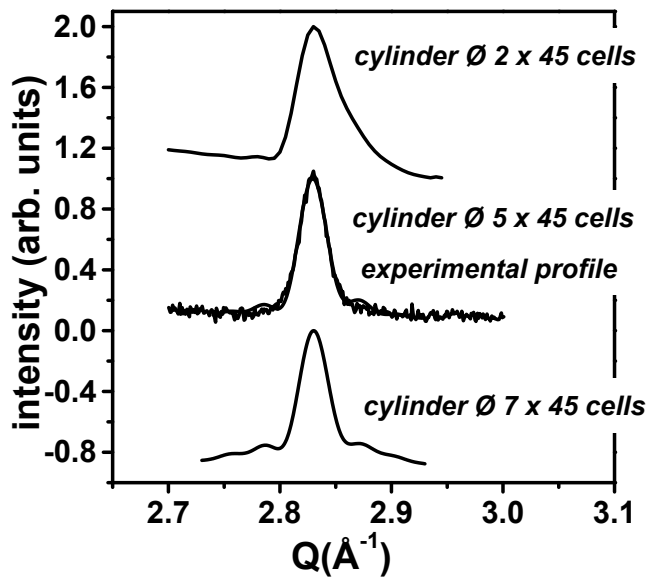}
\caption{Simulated profiles of the \{200\} reflection for cylinders of different diameters. At the center, the
observed profile of MnO in MCM matrix with 35 {\AA} channel diameter is shown for comparison.}
\label{profiles-1}
\end{figure}

Assuming that the cylinder axis is oriented along a [001] axis, the three contributions to the \{200\} peak
correspond to different apparent sizes. The width of the 200 and 020 reflections yields an apparent size
corresponding to the diameter, whereas for the reflection 002 the apparent size corresponds to the cylinder
height. Because in our case the diameter is much smaller than the height, the resulting lineshape is
practically defined by the narrowest line 002. Therefore the lineshape of the \{200\} peak measured for MnO
confined in MCM with 35 {\AA} channel diameter is well described by the analytical formula of Langford
developed for powder diffraction from cylindrical objects \cite{Langford} (figure \ref{profiles-3}). However
this formula does not work well for very narrow cylinders because it does not take into account the real
lattice. In figure \ref{profiles-1} we compare the experimental lineshape with that simulated by the Debye
formula.

Two effects are seen. Firstly, for the very narrow cylinders the calculated profile shows a "sawtooth" profile
as for thin layers. This is the expected result since the powder averaging in the reciprocal space is the same
for a thin layer and thin cylinder. Secondly, for the largest diameters there is a well defined broad
"pedestal" defined by the cylinder diameter, while the narrow peak is mainly defined by the cylinder height.
Neither a "sawtooth" profile, nor a broad "pedestal" were observed in the experiment with MCM. In figure
\ref{profiles-1}, the experimental profile measured for MCM with 35 {\AA} diameter was best matched by a
cylinder diameter of 5 cells (22 {\AA}).

Surprisingly, the diameter of the embedded nanoparticle of 22(3) {\AA} turns out to be significantly smaller
than the channel diameter of 35 {\AA}. It appears that the crystallization occurs only in the center of the
channel. There are two possible reasons: the roughness of the silica walls, evaluated to a few
{\AA}ngstr\"{o}ms or/and a boundary layer of amorphous manganese oxide. The marked difference in the diffuse
backgrounds for MCM and SBA (figure \ref{experiment}), which does not coincide with the diffuse background of
the unfilled matrices confirms the presence of such an amorphous phase.

A similar effect was noticed in previous studies of liquid oxygen confined to channel matrices \cite{Oxygen}.
The separation of confined liquid in two phases, an amorphous coating of the pore walls and a capillary
condensate in the channel center, was claimed. We possibly deal with a similar effect, since MnO crystallizes
in the channels from a liquid phase. To be more precise, one should investigate the process of transformation
of MnO at the interface versus external parameters. Additional oxidation processes, which depend on the type of
matrix (SBA/MCM), the conditions of preparation and conservation of the samples, clearly determine the
respective amounts of the amorphous and crystalline phase.

\section{Conclusion}

X-ray synchrotron and neutron diffraction experiments show that MnO crystallizes inside nanochannels of
mesoporous silica matrices adopting the shape of nanoribbons and nanowires. It leads to a complex lineshape and
a specific shift of the diffraction reflections ("Warren shift"). The experimentally measured Warren shift and
observed profile lineshape were compared with those calculated for different nanostructured objects, on the
basis of the Debye formula. This allowed us to estimate the dimensions of the diffracting objects.

In SBA matrices, MnO crystallizes as narrow and thin ribbons of (47-87 {\AA} width, 4-9 {\AA} thickness),
with lengths of 270-350 {\AA}. These ribbons might be slightly curved. In contrast, in MCM matrices MnO
crystallizes as narrow cylinders (nanowires) with a diameter of about 20 {\AA}, smaller than the channel
diameters, and a length of 180-200 {\AA}.

The small difference in the channel diameters of MCM and SBA (35 {\AA} and 47 {\AA} respectively for the
closest diameters) could hardly explain such different shapes of confined nanoparticles. Clearly, other
parameters like the nature of the wall surface and its wetting by the MnO play an important role in the
formation of the nanoparticle. These parameters, which are different in SBA and MCM, could be crucial to
determine the specific features (shape, dimension, orientation) of the particles.

\begin{acknowledgements}
The authors thank C. Alba-Simionesco, N. Brodie and G. Dosseh who prepared and characterized the MCM and SBA
matrices. They are very grateful to J. Rodriguez-Carvajal for a critical reading of the manuscript and for
fruitful discussions. They also thank D. Morineau, R. Almairac and S. Rols for useful discussions. The work was
supported by the RFBR (Grants 02-02-16981 and 04-02-16550) and the INTAS (Grant 2001-0826). One of us I.V.G.
acknowledges the financial support of C.N.R.S. during his stay in LLB.
\end{acknowledgements}

\end{document}